\documentclass[a4paper,11pt]{article}
\pdfoutput=1 % if your are submitting a pdflatex (i.e. if you have
\usepackage{jinstpub} % for details on the use of the package, please
\usepackage{caption}
\usepackage{subcaption}
\usepackage{lineno}
\title{End-to-end simulations of the MUon RAdiography of VESuvius experiment}

\author[a,1]{A. Samalan,\note{Corresponding author.}}
\author[b]{S. Basnet,}
\author[c,d]{L. Bonechi,}
\author[e,f]{L. Cimmino,}
\author[c,d]{R. D'Alessandro,}
\author[e,f]{M. D'Errico,}
\author[b]{A. Giammanco,}
\author[b]{R. Karnam,}
\author[g]{G. Macedonio,}
\author[b]{M. Moussawi,}
\author[a]{C. Rendon,}
\author[e,f]{G. Saracino,}
\author[a]{M. Tytgat,}

\affiliation[a]{Department of Physics and Astronomy, Ghent University,
Ghent, Belgium}
\affiliation[b]{Centre for Cosmology, Particle Physics and Phenomenology (CP3), Universit\'e catholique de Louvain,
Louvain-la-Neuve, Belgium}
\affiliation[c]{INFN sez. di Firenze, Florence, Italy}
\affiliation[d]{University of Florence, Florence, Italy}
\affiliation[e]{INFN sez. di Napoli, Naples, Italy}
\affiliation[f]{University of Naples Federico II, Naples, Italy}
\affiliation[g]{INGV, Osservatorio Vesuviano, Naples, Italy}

\emailAdd{amrutha.samalan@cern.ch}

\abstract{The MUon RAdiography of VESuvius (MURAVES) project aims at the study of the summital cone of Mt. Vesuvius, an active volcano near Naples (Italy), by measuring its density profile through muon flux attenuation. Its data, combined with those from gravimetric and seismic measurement campaigns, will be used for better defining the volcanic plug at the bottom of the crater.
We report on the development of an end-to-end simulation framework, in order to perform accurate investigations of the effects of the experimental constraints and to compare simulations, under various model hypotheses, with the actual observations. The detector simulation setup is developed using GEANT4 and a study of cosmic particle generators has been conducted to identify the most suitable one for our simulation framework. To mimic the real data, GEANT4 raw hits are converted to clusters through a simulated digitization: energy deposits are first summed per scintillator bar, and then converted to number of photoelectrons with a data-driven procedure. This is followed by the same clustering algorithm and same tracking code as in real data. We also report on the study of muon transport through rock using PUMAS and GEANT4.
In this paper we elaborate on the rationale for our technical choices, including trade-off between speed and accuracy. The developments reported here are of general interest in muon radiography and can be applied in similar cases.}

\keywords{Models and simulations, Simulation methods and programs, Particle tracking detectors}

\arxivnumber{2109.14324} % only if you have one

\collaboration[c]{on behalf of the MURAVES collaboration}

\proceeding{22nd International Workshop on Radiation Imaging Detectors\\
27 June 2021 to 1 July 2021}

\begin{document}
\maketitle
\flushbottom

\section{Introduction}

Muon radiography is an imaging technique that allows to study the interior of large scale natural and man-made objects by using the naturally occurring muons from cosmic showers~\cite{Bonechi:2019ckl}. This technique exploits the penetration capability of muons and the imaging is performed from the measurements of the absorption profiles of muons as they pass through matter. The MUon RAdiography of VESuvius (MURAVES) project~\cite{MURAVES} aims at the study of the summital cone of Mt. Vesuvius, an active volcano near Naples, Italy. The main expected outcome of MURAVES will be a ``muography'' of Mt. Vesuvius, i.e. a density profile derived from the attenuation of the muon flux as a function of the incoming direction of the muons. Muography can eventually be combined with gravimetry and seismotomography for better defining the volcanic plug at the bottom of the crater.

The detection setup of MURAVES consists of three identical and independent tracking hodoscopes, constituted of scintillator bars coupled to SiPM, which are already installed on Mt. Vesuvius and fully operational. Each hodoscope includes four tracking stations, with a thick lead wall between the 3rd and the 4th that acts as a passive filter for low-momentum muons. 

Comparing data with Monte Carlo (MC) simulations is crucial for the imaging of a target. 
A popular approach is to compare various density hypotheses with the observed transmission map, seeking the one providing the best fit. In other types of studies, one is interested in anomalies with respect to the expected map. In all those cases, various effects can bias the expectations, thus it is important to quantify their impact by testing various modeling assumptions in MC. 

The simulation chain starts from the generation of muons with realistic angular and momentum distributions. Methods range from {\it ab initio} simulations that simulate the entire cosmic shower in the atmosphere (including background particles such as electrons, positrons, hadrons), such as CORSIKA~\cite{CORSIKA}, to parametrizations based on existing data. 
At the receiving end of the cosmic cascade, the geometry and the response function of the detector itself need to be simulated, at different levels of detail depending on the case. While simple parametrizations of the detector efficiency are often employed in muography, the high statistical precision expected after some years of MURAVES data taking requires full simulations of the detector response. 
In between the source and the detector, the most crucial simulation step is the particle propagation through the target material, where absorption and scattering effects need to be accurately modeled. Multipurpose tools developed for particle and nuclear physics, such as GEANT4~\cite{GEANT4,GGS} or FLUKA~\cite{FLUKA}, provide accurate models of the muon interaction with the traversed material, but take a very long computation time at each simulation step.  
For this reason, parametric or semi-parametric simulations have been developed with the aim to reduce CPU time while preserving a fair accuracy (see e.g. \cite{MUSIC,MUSUN}).

\begin{figure}[htbp]
    \centering % \begin{center}/\end{center} takes some additional vertical space
    \includegraphics[width=\textwidth]{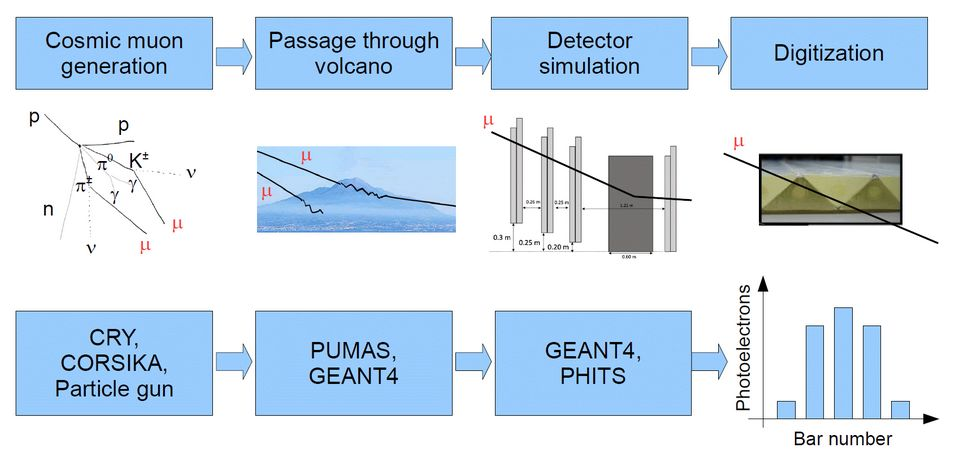}
    \caption{\label{fig:chain}A schematic representation of the MURAVES simulation chain.}
\end{figure}

In this paper, we report on the development of a MC simulation setup based on the schematic steps illustrated in Fig.~\ref{fig:chain}. 
We also present a series of simulation studies that are being conducted to investigate the effects of the experimental constraints and to perform comparisons with the actual observations. 
The detector simulation is based on GEANT4~\cite{GEANT4} and, for the generation of cosmic muons, a thorough comparison of CORSIKA~\cite{CORSIKA} and CRY~\cite{CRY} has been performed to identify the most suitable for our simulation framework. To mimic the real data, GEANT4 raw hits are passed through a simulated digitization, followed by the same clustering algorithm and thresholds as in real data. After application of the tracking algorithm, we compare a few high-level quantities with data.
For what concerns muon transport through rock, we are investigating PUMAS~\cite{PUMAS}, a ``backward MC'' designed to optimize computational efficiency, benchmarking it against GEANT4.

\section{Particle generation}

The first simulation step is the generation of muons, as well as background particles such as electrons, positrons, hadrons, according to a realistic model. 
We examined and compared two Monte Carlo tools with very different philosophy and complementary merits: CRY~\cite{CRY} and CORSIKA~\cite{CORSIKA}. 
CRY is a parametric single-particle simulation while CORSIKA performs a complete step-by-step development of a cosmic shower. 
There is a huge difference in their speed: we estimated that to generate 10$^5$ muons, CRY employs 1.4 minutes while CORSIKA takes 9 hours on a standard CPU. Therefore, they fulfill different use cases: CRY is intended to be used when speed is more relevant than accuracy, and CORSIKA the other way around. 

\begin{figure}[htbp]
\centering % \begin{center}/\end{center} takes some additional vertical space
\includegraphics[width=10cm, height=7cm]{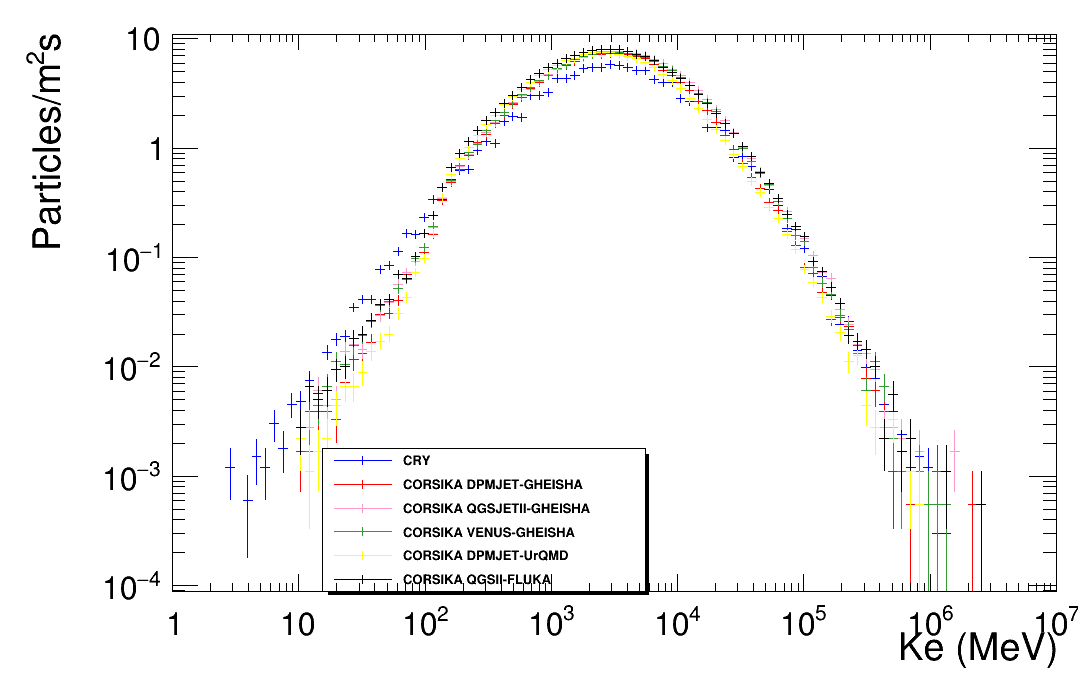}
\caption{\label{fig:spectrum}Muon energy spectrum using CRY and five hadronic interaction models of CORSIKA.}
\end{figure}

The two simulations complement each other in the possibility to address various systematic effects. CRY's parameterization, based on the MCNPX simulation~\cite{MCNPX}, takes into account geomagnetic effects on the cosmic flux, depending on time (solar cycle), latitude and altitude. On the other hand, in CORSIKA the user can choose across various low-energy and high-energy hadronic interaction models.
Figure \ref{fig:spectrum} shows the particle flux as a function of momentum, generated using CRY and five different combinations of high energy and low energy hadronic interaction models of CORSIKA. 
It can be seen that, the general shape and the peak positions are qualitatively similar for all models for the whole momentum range (1 MeV to 10$^7$ MeV). Given the dependence of both absorption and scattering on the particle momentum, these differences may lead to systematic uncertainties that need to be taken into account in the final measurement.

For the purposes of MURAVES data analysis, it was decided to use CRY as main generator, while also maintaining CORSIKA for the purpose of estimating systematic uncertainties related to physics modeling. 
The main reason for using CRY as default is its speed, but another practical reason is its ease of integration with GEANT4, in the detector simulation described in Sec.~\ref{sec:geant}. 
Integration of CORSIKA with GEANT4 is possible but much more complex. Considering also the slowness of CORSIKA and therefore the need to reuse its events as much as possible (while changing other parameters down the simulation chain), we save intermediate ascii files in HepMC format~\cite{HepMC} to then feed them to GEANT4.

%A fundamental limitation of CRY is that it cannot simulate more than one muon in the same event (not a very rare occurrence for large-area detectors), therefore we plan to use CORSIKA also to provide a realistic estimate of multi-muon events recorded by the MURAVES telescopes.
Although CRY is able to simulate more than one muon in the same event, CORSIKA is expected to provide a more realistic estimate of multi-muon events recorded by the MURAVES telescopes (not a very rare occurrence for large-area detectors).
Reliable simulated events with more than one muon will also help developing tracking algorithms that can cope with this occurrence (currently, in real-data analysis, two-muon events are either removed or a single-muon ansatz is imposed in the reconstruction.)

Other generators may be considered in the future; for example, a new parametric generator named EcoMug~\cite{EcoMug} has been recently released, whose guiding principle is to maximize computational efficiency, and which allows to shoot muons from a variety of generation surfaces (while CRY and CORSIKA are limited to flat surfaces). While these are appealing features, EcoMug does not simulate background particles, differently from both CRY and CORSIKA.

% CESAR

\section{Muon transport through Vesuvius}

\begin{figure}[htbp]
    \begin{minipage}[b]{0.40\textwidth}
        \includegraphics[width=\textwidth]{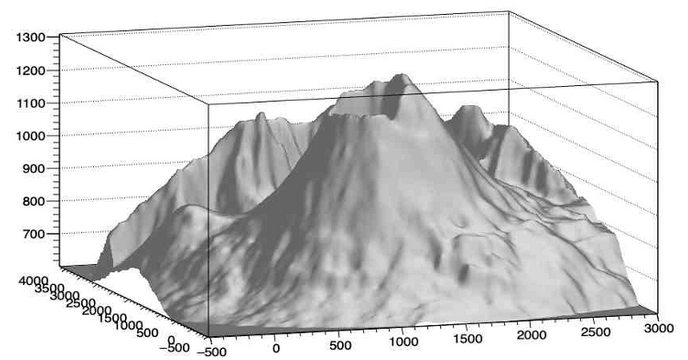}
        \caption{\label{fig:dtm}Digital terrain model of Vesuvius crater. Based on data from~\cite{vilardo2013morphometry}.}
    \end{minipage}
    \hfill
    \centering % \begin{center}/\end{center} takes some additional vertical space
    \begin{minipage}[b]{0.50\textwidth}
        \includegraphics[width=\textwidth]{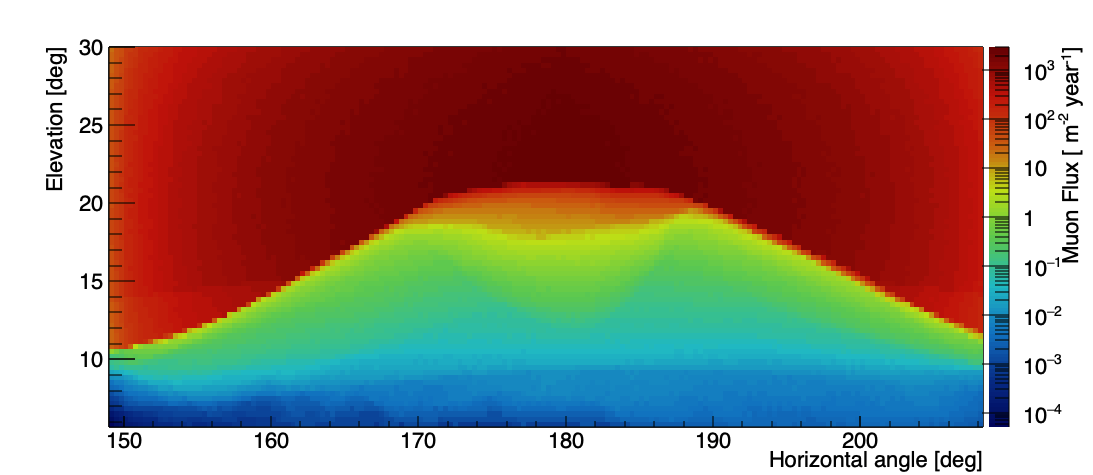}
        \caption{\label{fig:pumas}Expected muon flux per year considering the geometrical acceptance of the detector with PUMAS.}
    \end{minipage}
\end{figure}

The probability of muons to cross an object depends on the opacity of the latter ($\chi$, defined as the integral of the density along the path) and the initial energy of the muons. The measurement of the transmitted flux of atmospheric muons through a target from different directions provides the two-dimensional projective measurement of the target from the detector location and the average density ($\rho$) of the material of the target can be calculated by knowing the opacity using the equation $\chi=x\rho$, where $x$ is the muon path length or thickness of the material. The thickness distribution of the Vesuvius crater is evaluated using its Digital Terrain Model (DTM), derived using the data from~\cite{vilardo2013morphometry} shown in Fig. \ref{fig:dtm}.

% The measurement of the transmitted flux of atmospheric muons through a target allows inferring the integrated density along lines of sight. Dedicated Monte Carlo computations are required to precisely measure the density.
The simulation of the expected muon flux through Mt. Vesuvius (Fig. \ref{fig:pumas}) is conducted using PUMAS~\cite{PUMAS}, a muon transport library based on a Backward Monte Carlo (BMC) technique, integrated with TURTLE~\cite{TURTLE} for the navigation of the DTM. 
The BMC can be concisely described as ``running the MC backwards'', i.e. from the detector to the sky. 
The BMC deals with the exact same problem as the forward case and it allows to generate the final states that are actually observable, thus reducing the computation time considerably. In the case of our study, the scattering of the muons when they pass through the mountain was taken in to account by PUMAS and the muon flux was simulated given the thickness and average density of the material. In locations which are close to the mountain, we can expect considerable amount of low energy muons suffering multiple scattering near the surface and being deflected into the acceptance of the hodoscope at a fake direction. By accounting such background effects, the PUMAS simulation was also beneficial for the choice of the optimal observation point for MURAVES detector. 
% \begin{figure}[htbp]
% \centering % \begin{center}/\end{center} takes some additional vertical space
% \includegraphics[width=\textwidth]{figures/g4_pumas.png}
% \caption{\label{fig:geant-vs-pumas}Distributions of energy loss (a) and scattering angle (b) of 0-100 GeV muons in 50 m of standard rock generated using GEANT4 and PUMAS.}
% \end{figure}
\begin{figure}\centering
\subfloat[]{\label{a}\includegraphics[width=.49\linewidth]{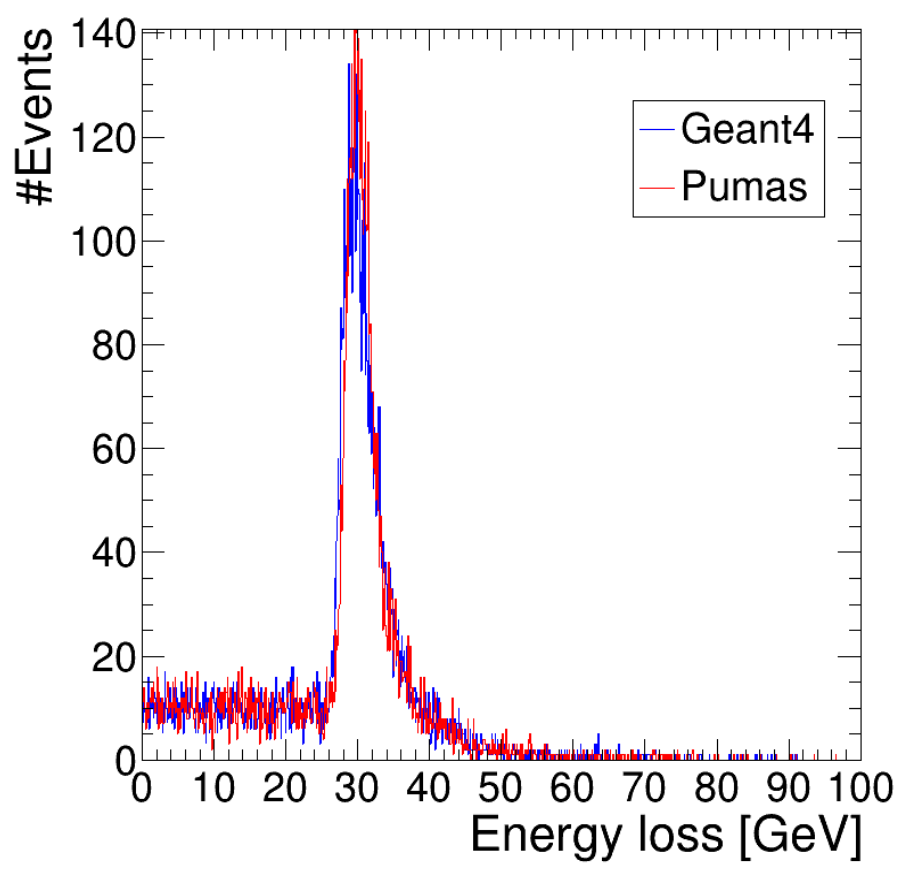}}\hfill
\subfloat[]{\label{b}\includegraphics[width=.49\linewidth]{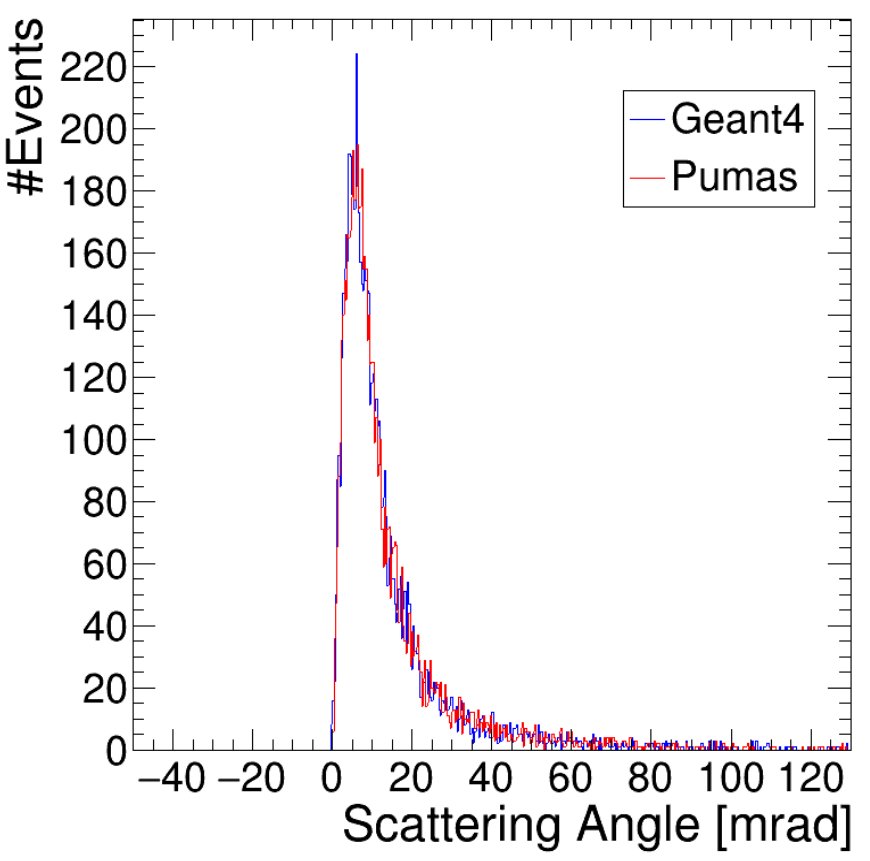}}\par 
\caption{Distributions of energy loss (\ref{a}) and scattering angle (\ref{b}) of 0-100 GeV muons in 50 m of standard rock generated using GEANT4 and PUMAS.}
\label{fig:geant-vs-pumas}
\end{figure}
Given the substantially larger computation efficiency of PUMAS with respect to forward MC programs, it was decided to use it as main program for this simulation step, but to validate it against a ``golden standard''. 
For that reason, we studied the energy loss and scattering of muons in the rock using PUMAS and GEANT4, and the results of the two simulations were compared. Muons in the energy range from 0 GeV to 100 GeV were simulated and the muon path length was set to 50 m in standard rock. Figure \ref{fig:geant-vs-pumas} shows the distribution of energy loss (\ref{a}) and scattering angle (\ref{b}) evaluated using PUMAS (red line) and GEANT4 (blue line) and both methods produced almost same results.

% Figure \ref{fig:geant-vs-pumas} shows the results of a comparison between PUMAS and Geant4~\cite{GEANT4} for the study of the energy loss (a) and scattering angle (b) of 0-100 GeV muons in 50 m of standard rock.

% \begin{figure}[htbp]
% \centering % \begin{center}/\end{center} takes some additional vertical space
% \includegraphics[width=13cm, height=6cm]{figures/crater.png}
% \caption{\label{fig:dtm}Digital terrain model of Vesuvius crater. Data from~\cite{vilardo2013morphometry}.}
% \end{figure}

% \begin{figure}[htbp]
% \centering % \begin{center}/\end{center} takes some additional vertical space
% \includegraphics[width=\textwidth]{figures/pumas1.png}
% \caption{\label{fig:pumas}Expected muon flux considering also the geometrical acceptance of the detector with PUMAS.}
% \end{figure}

\section{Detector response}
\label{sec:geant}

GEANT4~\cite{GEANT4} is the most popular MC particle transport simulation in particle and nuclear physics, as well as medical and space applications. 
For the MURAVES simulation studies, we built a detailed model of a MURAVES hodoscope in GEANT4, interfaced with CRY as source for all signal and background particles. For specific studies, single muons are simulated at specific energies and incident positions using a so called ``particle gun'', or the General Particle Source class.

\begin{figure}[htbp]
\centering % \begin{center}/\end{center} takes some additional vertical space
\includegraphics[width=10cm, height=4cm]{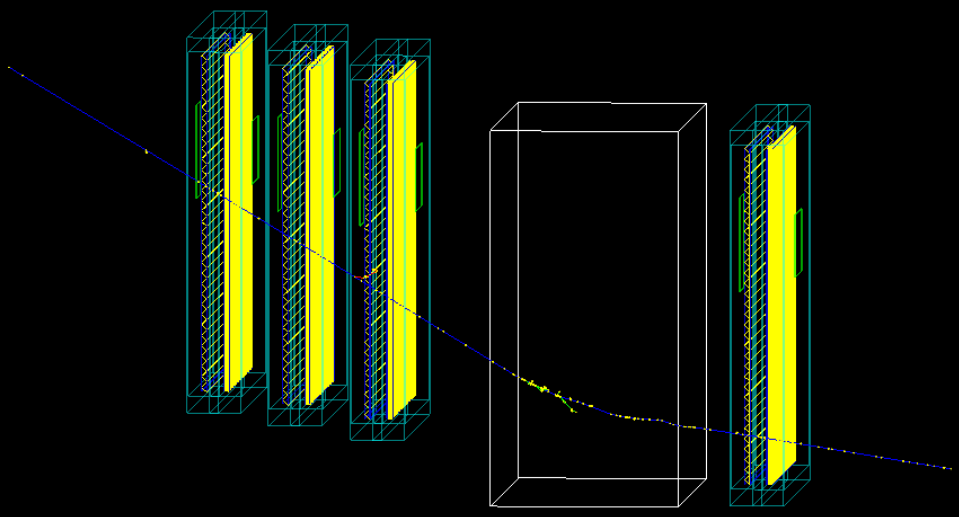}
\caption{\label{fig:1gev}Interaction of a 1 GeV muon with one of the the MURAVES hodoscopes in GEANT4.}
\end{figure}

The MURAVES setup consists of an array of three identical and independent tracking hodoscopes, constituted of scintillator bars coupled to SiPM, which are already installed on Mt. Vesuvius and fully operational. 
%In our MC simulation we do not distinguish between the three.
The MC simulation is applied to the three hodoscopes, considered as strictly identical.
Each hodoscope includes four tracking stations, with a thick lead wall of thickness 60 cm between the 3rd and the 4th that acts as a passive filter for low-momentum muons. Each station has two planes (X and Y) with 64 adjacent scintillator bars and of 1 m$^2$ sensitive area placed orthogonal to each other to extract the bi-dimensional information of the muons hits. The scintillator bars are of triangular shape with base width of 33 mm and height 17 mm, and alternate upward and downward pointing tips. All these detector parameters are implemented in the simulation setup and Fig. \ref{fig:1gev} shows the interaction of a 1 GeV muon with a MURAVES hodoscope.

A study on the probability for a muon to reach the 4th station of the detector (after surviving the lead wall) was performed using the particle gun. This probability is defined as the ratio of the number of events that reached the 4th station to the number of events which crossed the first 3 stations. As shown in Fig. \ref{fig:turnon}, we found that 50$\%$ survival probability is reached at 0.9 GeV (dotted red line) and almost 100$\%$ at 1.2 GeV (solid green line).

Another program, PHITS~\cite{PHITS} (interfaced to PARMA/EXPACS~\cite{sato2015analytical} as muon generator), was considered as a potentially valid alternative to GEANT4, and we succeeded to build an approximate representation of the MURAVES setup also in that framework. However, not all geometric details could be easily simulated, hence we chose to focus on GEANT4 for the detector response simulation, although this choice might in principle be reconsidered in the future.

\section{Digitization, clustering and tracking}

%The GEANT4 output data are processed such to mimic the real data generated by the MURAVES detectors. 

The output of GEANT4 contains ``hits'', corresponding to each simulation step for each particle (e.g., a hit may correspond to the interaction of a charged particle that leads to the emission of a photon). Hits are characterized by 3D position and time, energy deposition, and identity of the particle involved. 
In GEANT4, when a muon passes through a detector plane, it typically produces multiple hits in the scintillator bars, while we get only a single signal per bar in the real case. 

Our next simulation step is the ``digitization'' of these GEANT4 raw hits, i.e. the quantisation of both their positions and energy deposits. 
Energy deposits are first summed per scintillator bar, and then converted to number of photoelectrons with a data-driven procedure.
%{\bf Request to Samip: describe this procedure.}
At this point we have a single value of the energy deposit per bar, and we can apply clustering. A cluster is a collection of adjacent bars whose signals pass a certain threshold; clusters may also be constituted of a single bar, in which case however the signal must pass a stricter threshold. 
When a cluster contains more than one bar, the cluster position is calculated from their barycenter, i.e. the weighted average of the positions of the bars involved, where the weights are given by the number of photoelectrons. 
The thresholds to be used are still rapidly evolving; here it suffices to say that rigorously the same algorithm and thresholds are applied in MC as in real data at any point in time. 

\begin{figure}[htbp]

    \begin{minipage}[b]{0.49\textwidth}
        \includegraphics[width=\textwidth]{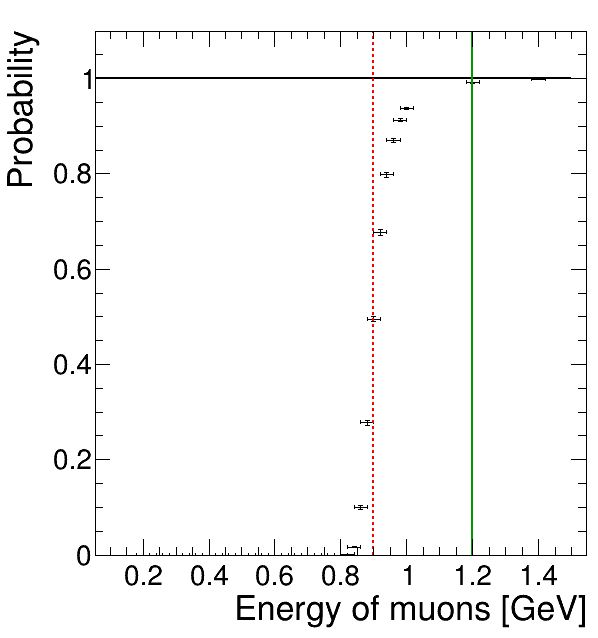}
        \caption{\label{fig:turnon}Probability for a muon to survive the passage through the lead wall, as a function of its energy.}
    \end{minipage}
    \hfill
    \centering % \begin{center}/\end{center} takes some additional vertical space
    \begin{minipage}[b]{0.49\textwidth}
        \includegraphics[width=\textwidth]{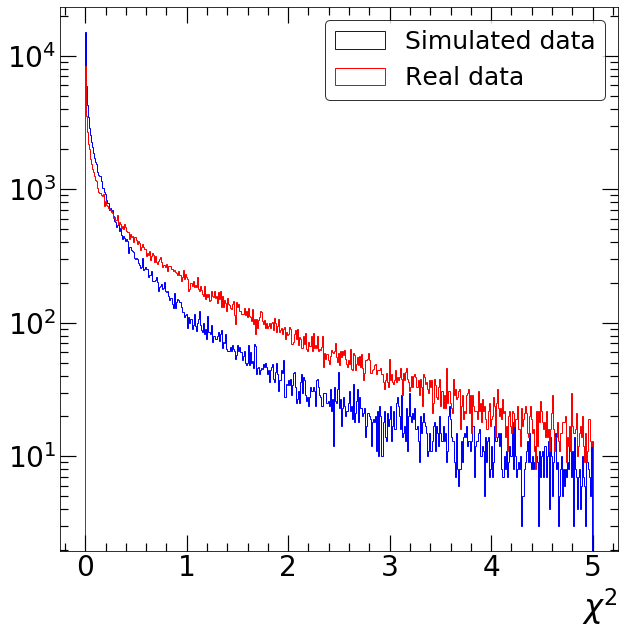}
        \caption{\label{fig:chi2}Comparison of $\chi^2$ of the tracks in real data and simulated data, both in free-sky conditions.}
    \end{minipage}
\end{figure}

Finally, the simulated clusters are fed into the tracking algorithm, which again is rigorously applied as in real data. 
The tracking is performed for X planes and Y planes independently. 
To validate the full simulation chain, the $\chi^2$ values of the simulated tracks were compared with those of the real data from a free-sky calibration run.
For this, 10$^5$ muon events were simulated, all crossing at least three planes of the detector, and the full reconstruction chain was applied; a $\chi^2 < 5$ selection is applied to the track sample. 
The result is shown in Fig. \ref{fig:chi2}. %The red line shows the distribution of the $\chi^2$ of simulated tracks and the blue line shows that of the real data. 
Although there is a slight variation in the tail, whose main causes are under investigation, the trend is qualitatively similar. This is far from being a trivial achievement for a simulation that does not consider (yet) several nuisances, such as dark noise, signal thresholds and various front-end electronics effects.

\section{Conclusions}

% We presented a complete Monte Carlo simulation chain for the MURAVES experiment. 
% For each step we examined some of the available software tools, striving for a trade-off between accuracy and speed that adequately meets the requirements of an experiment such as MURAVES.

% For the generation of the cosmic muons, we chose CRY as our workhorse, but CORSIKA finds a role due to the possibility to simulate additional effects (e.g. multiple muons in the same event) and to quantify the effect of different hadronic interaction models.
% The problem of efficient simulation of the muon passage through a large body, such as Vesuvius, is approached using PUMAS, whose accuracy we verified by comparing with GEANT4 in reference conditions. 
% We interface CRY with GEANT4 for the simulation of the detector response, and the output of GEANT4 is then digitized into a raw-data format that can be directly fed into the standard MURAVES reconstruction code. 
We presented a complete Monte Carlo simulation chain for the MURAVES experiment. 
For each step we examined some of the available software tools, striving for a trade-off between accuracy and speed that adequately meets the requirements of an experiment such as MURAVES.

For the generation of the cosmic muons, we chose CRY as our workhorse, but CORSIKA finds a role due to the possibility to simulate additional effects (e.g. multiple muons in the same event) and to quantify the effect of different hadronic interaction models.
The problem of efficient simulation of the muon passage through a large body, such as Vesuvius, is approached using PUMAS. To verify the accuracy of this method, we conducted a comparison study between PUMAS and Geant4 by analyzing the energy loss and scattering of muons in rock. The results of both methods were in line with each other. The survival probability of the muons in the MURAVES setup was also studied and as per the results, 50$\%$ of muons of energy 0.9 GeV were able to reach the 4th station of the detector and $\sim$100$\%$ survival probability was observed at 1.2 GeV. We interface CRY with GEANT4 for the simulation of the detector response, and the output of GEANT4 is then digitized into a raw-data format that can be directly fed into the standard MURAVES reconstruction code. To substantiate the performance of the simulation chain, we mimicked the tracking method applied for the real data for the simulation, and compared the $\chi^2$ values of the simulated tracks with those of the real data. The trend observed in both cases are similar but a slight difference is present due to not taking account of the experimental effects such as dark noise, electronics effects etc. in the simulation. 

\acknowledgments

The authors gratefully acknowledge the precious help by Valentin Niess and Cristina Carloganu; their simulations of Mt. Vesuvius with PUMAS~\cite{PUMAS} were used at the start of the MURAVES project for the choice of the optimal location for the installation of the detectors. 
We are also indebted to Pasquale Noli and Nicola Mori for their previous work on the simulation of MU-RAY (precursor of MURAVES)~\cite{GGS}. 
Paolo Strolin provided useful feedback on an early draft of this paper. 
This work was partially supported by the EU Horizon 2020 Research and Innovation Programme under the Marie Sklodowska-Curie Grant Agreement No. 822185, and by the Fonds de la Recherche Scientifique - FNRS under Grants No. T.0099.19 and J.0070.21. %Samip Basnet and Raveendrababu Karnam would like to acknowledge additional research grants from FNRS (FRIA doctoral grant) and UCLouvain (FSR postdoctoral fellowship), respectively.


%\begin{thebibliography}{99}

%\bibitem{a}
%Author, \emph{Title}, \emph{J. Abbrev.} {\bf vol} (year) pg.
%
%\bibitem{b}
%Author, \emph{Title},
%arxiv:1234.5678.
%
%\bibitem{c}
%Author, \emph{Title},
%Publisher (year).

% Please avoid comments such as "For a review'', "For some examples",
% "and references therein" or move them in the text. In general,
% please leave only references in the bibliography and move all
% accessory text in footnotes.

% Also, please have only one work for each \bibitem.

%\end{thebibliography}

\bibliographystyle{unsrt}
\bibliography{muraves}

\end{document}